\documentstyle[11pt,aas2pp4]{article}

\begin{document}
\slugcomment{Accepted to the Astrophysical Journal}

\title{Distances to the high galactic latitude molecular clouds \\ G192-67 and MBM~23-24}
\author{Catherine E. Grant\altaffilmark{1} and David N. Burrows\altaffilmark{1}}
\affil{Department of Astronomy and Astrophysics, Pennsylvania State University, \\ 525 Davey Lab, University Park, PA 16802}
\authoremail{cgrant@astro.psu.edu,burrows@astro.psu.edu}

\altaffiltext{1}{Visiting Astronomer, Kitt Peak National Observatory, National Optical Astronomy Observatories, which is operated by the Association of Universities for Research in Astronomy, Inc.\ (AURA) under cooperative agreement with the National Science Foundation.}

\begin{abstract}

We report on distance determinations for two high Galactic latitude cloud complexes, 
G192-67 and MBM~23-24.  No distance determination exists in the literature for
either cloud.  
Thirty-four early type stars were observed towards the two clouds, more than 
half of which have parallaxes measured by the {\it Hipparcos} satellite.  For the 
remaining stars we have made spectroscopic distance estimates.  The data consist of 
high resolution echelle spectra centered on the \ion{Na}{1}~D lines, and were obtained over 
six nights at the Coud\'{e} Feed telescope at Kitt Peak National Observatory.  
Interstellar absorption lines were detected towards some of the stars, 
enabling estimates of the distances to the clouds of $109 \pm 14$~pc for G192-67, and of $139 \pm 33$~pc for MBM~23-24.  We discuss the relationship of these clouds to other ISM features such as the Local Hot Bubble and the local cavity in neutral hydrogen.

\end{abstract}


\twocolumn

\section{Introduction}

We report on a search for interstellar \ion{Na}{1}~D lines towards two molecular cloud complexes, G192-67 and MBM~23-24.  These are low-mass, translucent clouds with molecular cores which do not appear to be forming stars, and are part of a survey of X-ray shadows towards molecular and cometary clouds using the {\it ROSAT} PSPC (\cite{gra98}).  Based on these measurements, we determine the distance to both clouds for the first time.

G192-67 ($\ell,b \sim 192 \arcdeg, -67 \arcdeg$) was discovered by \cite{oden87} in a search for comet-like clouds in the 100~$\micron$ band of the {\it IRAS} all-sky survey. \cite{oden88} found no evidence of optical nebulosity or depressed star counts from obscuration towards this cloud.  Neutral hydrogen 21-cm measurements of the cloud core (the `head' of the cometary formation) were carried out by \cite{hrk88}.  They found multiple velocity components at $V_{LSR}$ = 3.2, -9.2, and 0.9~km/s and a total \ion{H}{1} column density of $1.25 \times 10^{20}$~cm$^{-2}$.  This cloud was also detected in the molecular CO(1-0) line by \cite{rkh94} at $V_{LSR}$ = -3.8~km/s.

The MBM~23-24 cloud complex ($\ell,b \sim 172 \arcdeg, +27 \arcdeg$) was discovered as part of a survey of high-latitude molecular gas (\cite{mbm85}).  It is described as a small, fragmented complex, with two cores at $V_{LSR}$ = -2.6 and -1.5~km/s.  Little or no star formation is taking place, as evidenced by the lack of bipolar outflows (\cite{va90}) and absence of H$\alpha$ emitting stars (\cite{kun92}).  Neutral hydrogen measurements by \cite{gbm94} yielded velocities of $V_{LSR}$ = -3.3 and -3.6~km/s and an average \ion{H}{1} column density of $2.9 \times 10^{20}$~cm$^{-2}$.  Examination of \cite{gbm94} Figures 2 and 3d shows that the velocity structure is asymmetric with many small peaks between -10 and +15~km/s.

Many methods can be used to determine distances to interstellar clouds. These clouds have small radial velocities because of their positions (near the Galactic anticenter and high latitude) and probable small distances, and are not suitable for accurate kinematic distance determinations.  Other methods rely on the absorbing effect of the cloud on background stars.  If the cloud is thick enough and the stellar density distribution is understood, the deficit of stars toward the cloud can be used to calculate distance.  The accuracy of this technique is dependent on the size of the deficit, which is small for these types of clouds.  Comparing the stellar distance versus reddening of stars toward the cloud can yield a distance, but requires accurate spectral types in order to determine extinction.  We have chosen to measure the narrow absorption lines of \ion{Na}{1}~D  in optical spectra of stars towards the clouds.  The presence of an absorption line requires the star to be behind the cloud, while the lack of an absorption line implies the reverse.  Since these clouds are at high latitudes, few interloping absorption features are expected.  This method requires both high resolution spectroscopy and accurate stellar distances.  For stars observed by ESA's {\it Hipparcos} mission, stellar distances and their associated errors are accurately known without need for precise photometric data or spectral types.  Section~2 describes the selection of candidate stars and the high resolution spectroscopic measurements, \S~3, the assignment of distances to the clouds themselves, and \S~4 discusses the relation of these clouds to other interstellar features.

\section{Stellar data}

\begin{figure}
\plotone{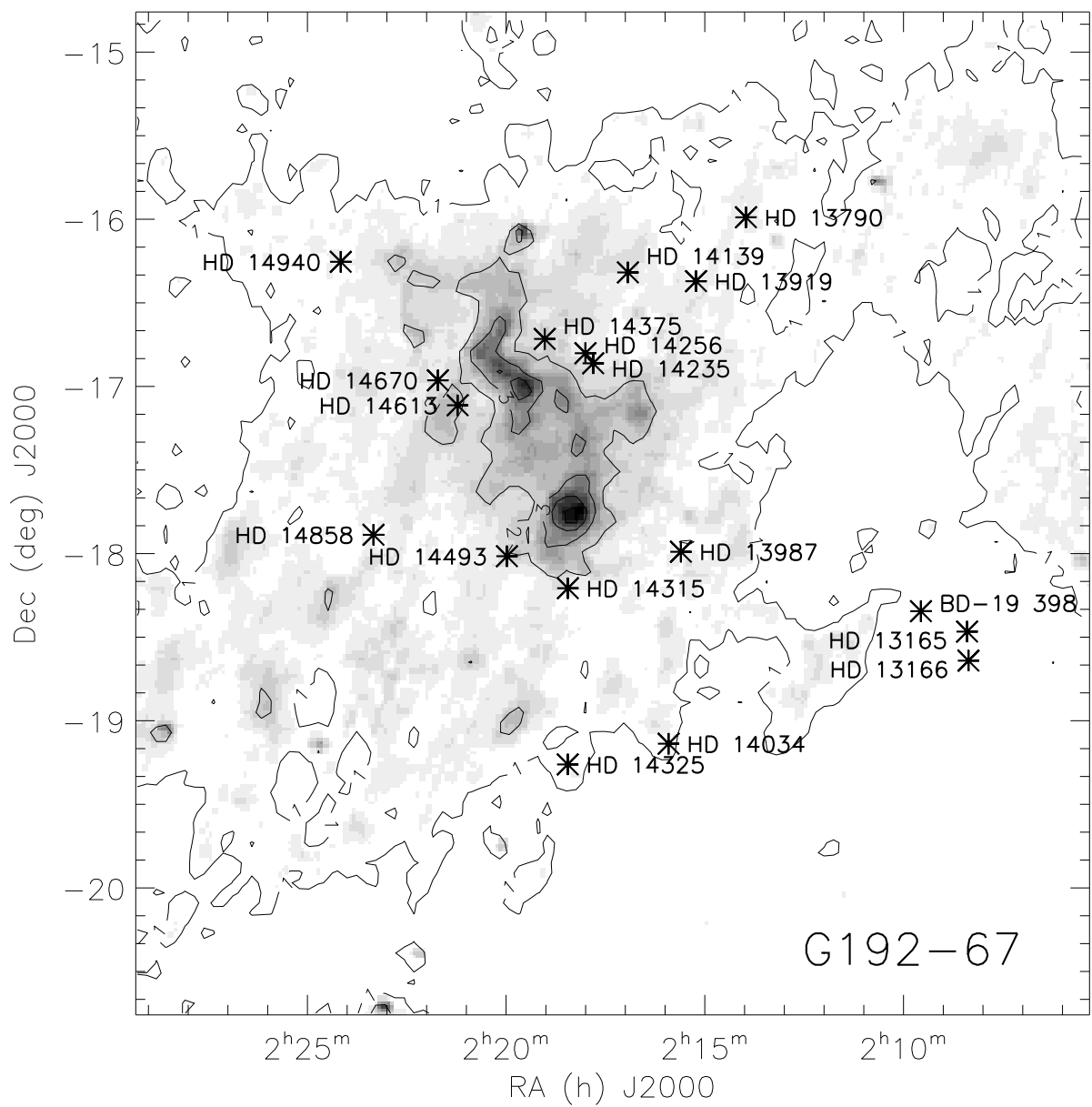}
\figcaption[]{{\it IRAS} 100 $\micron$ images of the molecular clouds.  Stars observed are labeled by asterices.  Contours are labeled and are separated by 1 MJy/sr.  a) G192-67 \label{fig-image}}
\end{figure}

Thirty-four stars were selected from the SIMBAD database using the following criteria: positioned within the infrared contours of the cloud, $V$ magnitude brighter than 10 and spectral type earlier than F8.  The selected objects are listed in Table~\ref{tbl-stardata} and their positions are shown in Figure~\ref{fig-image}.  Spectral types and $V$ magnitudes are from SIMBAD except where more precise magnitudes exist in the SKY2000 catalog (\cite{sky}).  $I_{100}$ is the surface brightness from the {\it IRAS} survey in the 100~$\micron$ band at the star's position.

Late F stars are not ideally suited for interstellar absorption studies since the stellar lines of \ion{Na}{1} are stronger and less rotationally broadened than earlier type stars.  We included these, however, in the hope that some would have stellar radial velocities with enough separation from the cloud velocity to permit observation of the interstellar lines. 

\subsection{Stellar Distances}

Primary stellar distances are from the {\it Hipparcos} catalog of stellar parallaxes ($D_H$).  In addition, we have calculated spectroscopic distances both excluding ($D_0$) and including ($D_R$) corrections for interstellar extinction, i.e. $D_0=10^{(V-M_V+5)/5}$ and $D_R=10^{(V-M_V+5-A_V)/5}$.  Stars without full MK spectral types are assumed to be on the main sequence with luminosity class V.  Absolute magnitudes and colors are from \cite{lang92}.  Since small errors in magnitudes can yield large errors in extinction-corrected distance, and since $A_V$ should be small, we are using only $D_0$ in later discussions.  When possible, our results are based solely on the more accurate {\it Hipparcos} distances; however those objects without {\it Hipparcos} measurements will still be considered.

To evaluate the reliability of our spectroscopic distances, we have made a statistical comparison between the spectroscopic and parallactic distances for those stars with {\it Hipparcos} measurements.  As seen in Figure~\ref{fig-dvd}, there is significant scatter which increases with $D_H$.  The rms deviation is 38\%, however the differences are not random, with $D_0$ underestimating the true distance, $D_H$.  The assumption of luminosity class V for all stars without full spectral types may be in error and a more evolved spectral type would significantly increase $D_0$.  Adding extinction effects to the spectroscopic distance would decrease the derived distance and in most cases appears to be unnecessary.  It is beyond the scope of this paper to examine in detail the errors which cause this discrepancy, but better photometry and spectral typing would most likely improve the results.

\cite{pin98} have examined apparent discrepancies between {\it Hipparcos} and main-sequence fitting distances to some open clusters, in particular the Pleiades in which the distances disagree by more than 3$\sigma$.  They conclude that systematic errors exist in the Pleiades parallax data at the 1 mas level.  Two possible sources of this error are spatial correlations on small angular scales ($\sim 1\arcdeg$) and statistical correlations among the fitted parameters (position, parallax and proper motion).  While our fields are not as compact as the Pleiades, some of the individual parallax measurements are not completely independent, having been observed in the same great circle scans.  The statistical correlation between parallax and right ascension ($\rho^\pi_{\alpha}$, field H20 in the {\it Hipparcos} catalog) is caused by uneven distribution of observations over the parallactic ellipse and is a strong effect only in certain areas of the sky (for the Pleiades $< \rho^\pi_{\alpha} > = 0.34$).  This correlation is small towards G192-67 ($< \rho^\pi_{\alpha} > = 0.02$), but unfortunately is large towards MBM~23-24 ($< \rho^\pi_{\alpha} > = 0.37$).  If the same kind of systematic error exists for these fields as does for the Pleiades, it would, at most, change our cloud distance estimates by 10-15 pc, which is within our estimated 1$\sigma$ error bars. 

\subsection{High-Resolution Spectroscopy}

We observed the program stars on 1997~November~18-23 at the 0.9-m Coud\'{e} Feed telescope at Kitt Peak National Observatory.  The echelle grating and cross-dispersing grism were used in the Camera 5 configuration, yielding a spectral resolution of R $\gtrsim$ 60,000.  Exposures of a thorium-argon lamp were taken every few hours as a wavelength reference.  The FWHM of the Th-Ar spectrum near the \ion{Na}{1}~D lines was 0.072~\AA\@  or 3.7~km/s.  The slit width was 300~$\micron$  or about 2 arcsec.  Total exposure times depended on magnitude and ranged from 1200 to 14,000 seconds.  No single exposure was longer than forty minutes to minimize contamination by cosmic rays.  Atmospheric transparency was poorer in the early evening, therefore the signal-to-noise achieved for stars in the G192-67 field is worse.  The mean S/N for the G192-67 field was $\sim$ 40 and for the MBM~23-24 field was $\sim$ 55.

The data were processed using the standard IRAF data reduction package (\cite{iraf}).  The raw CCD frames were bias-subtracted and flat-fielded, and bad pixels and cosmic rays were removed.  The echelle orders were extracted, sky-subtracted, and dispersion-corrected using the Th-Ar spectrum as a wavelength reference.  The stellar continua were normalized using a high order spline, and velocity shifted to the local standard of rest.  High signal-to-noise observations of the stars $\rho$~Ceti (A0V, $V_{rot}\sin i \approx$ 200 km/s) and 2~Ceti (B9IVn, $V_{rot}\sin i \approx$ 180 km/s) were combined to produce a typical telluric spectrum.  Atmospheric telluric absorption lines were removed by scaling the telluric spectrum to match the stellar spectrum (see \cite{lil90} for more details).  The normalized stellar spectrum was then divided by the scaled telluric spectrum.

The spectra for \ion{Na}{1}~D2 are shown in Figure~\ref{fig-spec}.  The radial velocity, equivalent width, and column density of the interstellar lines detected are shown in Table~\ref{tbl-ew}.  Absorption features were required to be present in both D1 and D2 spectra at similar velocities.  A multi-line entry indicates that multiple velocity components are present.  Column densities were calculated using N$_m ($cm$^{-2}) = 1.13 \times 10^{20} W_\lambda / \lambda_0^2 f_{mn}$ where $W_\lambda$ is the equivalent width in \AA, $\lambda_0$ is the rest wavelength of the absorption line and $f_{mn}$ is the oscillator strength of the transition.  This linear relation assumes that the line is optically thin and unsaturated.  To reduce the effects of saturation on the column density determination, we followed the empirically determined formulation of \cite{pen93}.  For $W_\lambda$(D2) $<$ 40 m\AA, N(\ion{Na}{1}) was calculated from the stronger D2 line.  For 40 m\AA $<$ $W_\lambda$(D2) $<$ 80 m\AA, N(\ion{Na}{1}) was calculated from the weaker D1 line.  For $W_\lambda$(D2) $>$ 80 m\AA, upper limits on N(\ion{Na}{1}) were calculated from the D1 line.  When a narrow stellar line occurs close to the expected velocity of the interstellar line, it is nearly impossible to detect the interstellar absorption.  These cases are ignored in further analysis.

\section{Cloud Distances}

Comments on individual stellar observations are made below.  Both clouds exhibit 21-cm emission over a range of velocities with multiple peaks.  All the suspected interstellar absorption components presented here are consistent with the velocities measured in neutral hydrogen.  For the MBM~23-24 observations which have multiple interstellar velocity components, we have assumed all are associated with the cloud.  We have also assumed that the N(\ion{Na}{1})/$I_{100}$ ratio is reasonably constant throughout the clouds so that stars with similar dust emission should show similar \ion{Na}{1} absorption.

\subsection{G192-67}

\paragraph{HD 14139}  
No interstellar absorption is seen in this spectrum which places a lower limit on the distance to G192-67 of $D_H = 101^{+16}_{-12}$~pc.  The infrared dust emission from the cloud at this point is 1.59 MJy/sr, similar to that of neighboring stars, HD~13790 (1.49), HD~14256 (1.45), and HD~14235 (1.69), which do show \ion{Na}{1} absorption.  It is possible that a chance juxtaposition of this star with a hole through the cloud could cause the lack of absorption, however the dust emission does not show this and it seems unlikely that the dust to gas ratio could change so drastically over such a short distance.

\paragraph{HD 14940}  
The only other star closer to us than HD~14139 is HD~14940 which is at $D_H = 62^{+3}_{-3}$~pc.  This star also shows no interstellar absorption which agrees with the previous result.

\paragraph{HD 14670}  
Interstellar absorption is seen at the velocity of the \ion{H}{1} emission in the spectrum of HD~14670, which is located at $D_H = 116^{+23}_{-17}$~pc.  This places an upper limit on the distance to the cloud.  More distant stars also show interstellar absorption, confirming this result.

\paragraph{HD 13166, HD 13165, BD-19 398}
These stars were added to the sample to confirm that the absorption seen in almost every spectrum is associated with the cloud and not with a diffuse sheet of interstellar gas.  These `off-cloud' observations of stars at distances of 225, 667, and 457~pc do not show any interstellar absorption at the cloud velocity, confirming the association.

\subsection{MBM 23-24}

\paragraph{HD 58520}
A lower limit can be placed on the distance to MBM~23-24 by the lack of absorption towards HD~58520, which is located at $D_H = 114^{+12}_{-10}$~pc.  Again the infrared dust emission in this direction is similar to that towards stars which do show interstellar absorption.

\paragraph{BD+47 1474}
The star BD+47 1474 shows strong interstellar absorption at the velocity of the \ion{H}{1} emission and is located at $D_H = 163^{+54}_{-33}$~pc.  This observation defines the best upper limit of the distance to MBM~23-24.

\paragraph{BD+47 1455}
The star BD+47 1455 has small but measurable interstellar absorption at the velocity of the \ion{H}{1} emission.  The distance is only known spectroscopically and is $D_0 = 120 \pm 46$~pc.  Since this star is located near to one of the cloud cores, either substantial or no absorption is expected.  If the star is behind the cloud, the small absorption line would require either a drastic change in the N(\ion{Na}{1})/$I_{100}$ ratio or a `hole' through the cloud that is unresolved in the {\it IRAS} data.  The star could be at the near edge of the cloud and absorbed by a small fraction of the total column density of the cloud; however no reflection of the star's light on the gas can be seen.  It seems most probable that the star is in front of the cloud and the absorption is due to the small amount of cold gas in the local cavity.  At this low column density, the expected extinction is vanishingly small, so that $D_0$ need not be corrected for reddening.  The published data do not include a luminosity class, so we have assumed that BD+47~1455 is on the main sequence.  A more evolved luminosity class of IV would increase the distance, to $D_0 = 209 \pm 79$~pc.  While the distance uncertainties are too great for this observation to constrain the cloud distance, there is an indication that the cloud may be further than 120~pc.

\paragraph{BD+47 1459}
Substantial interstellar absorption at the velocity of the \ion{H}{1} emission is observed toward BD+47~1459, which has a spectroscopically determined distance of $D_0 = 149 \pm 57$~pc.  This observation defines a lower, but less certain upper limit of the distance to MBM~23-24.

\subsection{Distance Estimates}
The derived distances to the clouds are as follows: for G192-67, $d = 109 \pm 14$~pc and for MBM~23-24, $d = 139 \pm 33$~pc.  These results are displayed in Figure~\ref{fig-dist} as a comparison of stellar distance versus total column density of interstellar \ion{Na}{1}.  For stars with multiple interstellar velocity components, the column density is the sum of these components.  A strong discontinuity is seen in both plots at the derived distance of the clouds. 

\section{Discussion}

The location of these clouds with respect to other large-scale structures in the ISM may be useful in understanding the origin and evolution of these kinds of low mass clouds.  The sun is embedded in the Local Hot Bubble (LHB), which is an amorphous region of hot (T $\sim 10^6$ K), rarefied ($n_H \lesssim 0.01 $ cm$^{-3}$) gas.  Very few molecular clouds are known to be within the LHB\@.  If more exist, these clouds could cool the hot gas very effectively, putting constraints on models of the formation and evolution of the LHB (\cite{lism}).  The exact size of the LHB is uncertain and model-dependent.  Using the middle scaling from \cite{lhb98}, the LHB extends to $\sim$ 85~pc towards G192-67 and $\sim$ 60~pc towards MBM~23-24.  In both cases, the clouds are outside the LHB\@.  The upper scaling from \cite{lhb98} ($\sim$~122~pc and $\sim$~86~pc, respectively) would allow G192-67 to lie at the edge or slightly inside the LHB, but MBM~23-24 would still remain far outside.

The local cavity of neutral hydrogen measured by the interstellar \ion{Na}{1}~D absorption line has a similar extent as the LHB but with different features (\cite{welsh94}, Snowden et al.\ 1998).  The data of Welsh et al.\ (1994), with {\it Hipparcos} distances, implies the edge of the local cavity may be further than the edge of the LHB in these directions, namely $\sim$~130~pc towards G192-67 and $\sim$~100~pc towards MBM~23-24; however this is based on a small sample of stars and is rather uncertain.  

It should be noted that the distribution of the denser gas is not always uniform, an example being a low-density tunnel towards $\beta$~CMa ($\ell \approx 235\arcdeg, b \approx -20\arcdeg$).  HD~13165 is located near G192-67, has a spectroscopically determined distance of $D_0 = 667$~pc, and an upper limit on \ion{Na}{1}~D column density of $0.7 \times 10^{11} $ cm$^{-2}$.  Using the empirical correlation between N(\ion{Na}{1}) and N$_H$ (\cite{fvg85}) this corresponds to N(\ion{H}{1}~+~H$_2$) $< 1.5 \times 10^{19} $ cm$^{-2}$.  This may imply that the local cavity extends much further in this direction or that its boundary is clumpy or patchy.  In high latitude regions, the cavity boundary may become more diffuse or disappear altogether.  More observations of high latitude stars are needed to investigate this.  

Placing these clouds at the interface between the LHB and the shell of the cavity is interesting theoretically, implying that these types of clouds may form as condensations as the LHB expands into the cooler, denser ISM (e.g. \cite{elm88}).  While the cloud distances do not match perfectly the extent of the LHB or local cavity, they are not wildly inconsistent.  The cometary appearance of G192-67 and the fragmentary appearance of both clouds suggest a more destructive process could be at work, with the LHB boundary shredding material off a pre-existing cloud inside the cavity (e.g. \cite{kmc94}).  As distances to more high latitude molecular clouds are measured and their association with large-scale ISM features understood, perhaps the answer will become apparent.

\acknowledgements
We would like to thank Bryan Penprase for help in preparing for our observing run and at the telescope.  We acknowledge the use of NASA's SkyView facility (http://skyview.gsfc.nasa.gov) located at NASA Goddard Space Flight Center.  This research has made use of the Simbad database, operated at CDS, Strasbourg, France.  This research was funded by a NASA Graduate Student Researchers Program fellowship (NGT5-50049).

\onecolumn
\clearpage

\begin{deluxetable}{lllcrcccr}
\small
\tablecolumns{9}
\tablewidth{0pt}
\tablecaption{Candidate Stars \label{tbl-stardata}}
\tablehead{
\colhead{Name} & \colhead{$\alpha_{ 2000}$} & \colhead{$\delta_{ 2000}$} & 
\colhead{Spectral} & \colhead{$V$} & \colhead{$I_{100}$} & \colhead {$D_{H}$} &
\colhead{$D_{O}$} & \colhead{$D_{R}$} \\
\colhead{} & \colhead{} & \colhead{} &
\colhead{Type} & \colhead{} & \colhead{(MJy/sr)} & \colhead{(pc)} &
\colhead{(pc)} & \colhead{(pc)} } 

\startdata

G192-67 \\
\cline{1-1} \\
 HD 13166     &2 08 21.91  &-18 38 26.1      &F7/F8V      &9.8   &0.30 &225     &160/144 &55/53 \\
 HD 13165     &2 08 24.00  &-18 28 00.0      &A0V         &9.8   &0.20 &\nodata &667     &506 \\
 BD-19 398    &2 09 33.72  &-18 20 40.8      &F8          &9.7   &0.50 &457     &140     &140 \\
 HD 13790     &2 13 57.88  &-15 59 17.4      &F5V         &8.9   &1.47 &\nodata &122     &122 \\
 HD 13919     &2 15 13.26  &-16 22 13.8      &F7V         &9.3   &1.33 &\nodata &128     &128 \\
 HD 13987     &2 15 36.07  &-17 59 20.9      &F5/F6V      &9.6   &1.19 &\nodata &167/160 &152/150 \\
 HD 14034     &2 15 54.71  &-19 08 14.3      &F3V         &8.8   &1.14 &201     &107     &92 \\
 HD 14139     &2 16 56.15  &-16 19 04.5      &F3V         &9.3   &1.59 &101     &134     &96 \\
 HD 14235     &2 17 48.50  &-16 51 42.5      &A8/A9V      &9.5   &1.69 &355     &264/252 &136/134 \\
 HD 14256     &2 18 00.22  &-16 48 04.5      &F5V         &9.1   &1.45 &142     &134     &134 \\
 HD 14315     &2 18 26.81  &-18 12 25.8      &F3V         &9.3   &1.41 &292     &134     &129 \\
 HD 14325     &2 18 26.96  &-19 15 40.3      &F7V         &9.3   &1.14 &164     &128     &46 \\
 HD 14375     &2 19 01.74  &-16 42 57.9      &F5V         &9.1   &1.94 &110     &134     &134 \\
 HD 14493     &2 19 58.74  &-18 00 56.8      &F2V         &9.0   &1.88 &160     &122     &117 \\
 HD 14613     &2 21 13.19  &-17 06 46.2      &F7V         &9.4   &2.35 &121     &134     &134 \\
 HD 14670     &2 21 43.03  &-16 57 48.1      &F6IV/V      &9.4   &1.65 &116     &233/147 &218/140 \\
 HD 14858     &2 23 20.25  &-17 53 15.6      &F5V         &9.4   &1.25 &\nodata &153     &78 \\
 HD 14940     &2 24 09.63  &-16 15 16.8      &F0IV/V      &7.1   &1.32 &62      &95/76   &95/76 \\ \\
MBM 23-24 \\
\cline{1-1} \\
 HD 58520     &7 28 08.65   &46 31 20.2      &A2          &6.9   &3.17 &114     &131     &130 \\
 HD 58549     &7 28 16.10   &46 17 13.7      &F5          &8.0   &3.46 &50      &78      &63 \\
 HD 59082     &7 30 28.89   &45 40 15.4      &F8          &8.9   &3.00 &103     &94      &94 \\
 BD+47 1455   &7 32 36.05   &47 09 31.4      &F2          &9.0   &3.79 &\nodata &120     &120 \\
 BD+47 1456   &7 32 45.62   &46 46 09.7      &F2          &9.3   &3.39 &124     &137     &106 \\
 BD+47 1459   &7 33 31.80   &47 19 25.9      &F5          &9.4   &5.31 &\nodata &149     &136 \\
 BD+46 1281   &7 33 50.71   &46 08 59.5      &F0          &8.9   &3.68 &\nodata &173     &110 \\
 HD 59975     &7 34 36.14   &48 11 28.8      &A3          &7.3   &3.44 &214     &144     &144 \\
 BD+47 1466   &7 36 57.75   &46 47 19.7      &F2          &9.1   &4.68 &\nodata &125\tablenotemark{a} &96\tablenotemark{a} \\
 HD 60653     &7 37 42.72   &45 03 33.6      &F2          &8.4   &2.61 &223     &90      &90 \\
 AG+47 701    &7 38 02.21   &47 29 55.9      &B8          &9.3   &3.52 &\nodata &813     &801 \\
 HD 60933     &7 39 03.48   &46 01 37.0      &A5          &8.7   &2.73 &\nodata &227     &227 \\
 BD+47 1470   &7 39 04.90   &47 11 58.0      &A0          &9.6   &3.30 &\nodata &608     &354 \\
 BD+46 1294   &7 40 08.72   &46 18 14.5      &A5          &9.5   &2.50 &\nodata &330     &214 \\
 BD+47 1474   &7 40 57.42   &46 59 11.0      &F8          &8.8   &2.64 &163     &92      &92 \\
 BD+47 1476   &7 41 15.15   &47 15 22.9      &F8          &9.9   &2.79 &165     &149     &93 \\
\enddata

\tablenotetext{a}{BD+47 1466 is a visual binary; these distances should be considered lower limits.}

\end{deluxetable}

\begin{deluxetable}{lrrrrrrc}
\small
\tablecolumns{8}
\tablewidth{0pt}
\tablecaption{Analysis of Stellar Spectra \label{tbl-ew}}
\tablehead{
\colhead{} & \multicolumn{5}{c}{Interstellar} & \colhead{Stellar} \\
\cline{2-6}
\colhead{Name} & \colhead{$V_{LSR}$} & \colhead{$W_{\lambda}$ (D1)} & 
\colhead{$W_{\lambda}$ (D2)} & \colhead{$\pm 2 \sigma$} & 
\colhead{N(\ion{Na}{1})} & 
\colhead{$V_{LSR}$} & \colhead{Notes} \\
\colhead{} & \colhead{(km/s)} & \colhead{(m\AA)} & \colhead{(m\AA)} & 
\colhead{(m\AA)} & \colhead{($\times 10^{11} $cm$^{-2}$)} & \colhead{(km/s)} & \colhead{}  } 

\startdata

G192-67 \\
\cline{1-1} \\ 

HD 13166   &\nodata &\nodata &\nodata &10 &\nodata  &-4.0    &a\\
HD 13165   &none    &$<$ 15  &$<$ 15  &15 &$<$ 0.7  &-3.1    & \\
BD-19 398  &\nodata &\nodata &\nodata &14 &\nodata  &-11.1   &a\\
HD 13790   &+1.4    &74      &51      &8  &7.4      &-10.8   & \\
HD 13919   &\nodata &\nodata &\nodata &9  &\nodata  &+6.5    &a\\
HD 13987   &+2.9    &97      &108     &10 &$>$ 9.7  &-25.8   & \\
HD 14034   &+4.9    &81      &110     &6  &$>$ 8.0  &-23.0   & \\
HD 14139   &none    &$<$ 10  &$<$ 10  &10 &$<$ 0.5  &-9.3    & \\
HD 14235   &+2.9    &92      &104     &10 &$>$ 9.1  &\nodata & \\
HD 14256   &+3.4    &20      &29      &9  &2.0      &-10.8   & \\
HD 14315   &+1.6    &82      &111     &8  &$>$ 8.1  &-13.6   & \\
HD 14325   &\nodata &\nodata &\nodata &7  &\nodata  &-13.4   &a\\
HD 14375   &\nodata &\nodata &\nodata &8  &\nodata  &-2.7    &a\\
HD 14493   &\nodata &\nodata &\nodata &8  &\nodata  &+16.2   &a\\
HD 14613   &+1.4    &122     &129     &9  &$>$ 12.2 &+33.2   & \\
HD 14670   &+0.6    &52      &55      &8  &5.1      &+14.4   & \\
HD 14858   &+1.9    &62      &60      &7  &6.2      &-4.0    & \\
HD 14940   &none    &$<$ 13  &$<$ 13  &13 &$<$ 0.6  &-11.1   & \\ \\

\tablebreak
MBM 23-24 \\
\cline{1-1} \\ 

HD 58520   &none    &$<$ 5   &$<$ 5   &5  &$<$ 0.5  &+27.6   & \\
HD 58549   &\nodata &\nodata &\nodata &5  &\nodata  &-24.0   &a\\
HD 59082   &\nodata &\nodata &\nodata &6  &\nodata  &+8.2    &a\\
BD+47 1455 &+8.5    &16      &20      &7  &1.0      &+3.9    & \\
BD+47 1456 &\nodata &\nodata &\nodata &8  &\nodata  &+4.7    &a\\
BD+47 1459 &+11.6   &86      &106     &7  &$>$ 8.6  &+8.2    & \\
BD+46 1281 &-6.2    &32      &38      &9  &1.9      &\nodata & \\
           &+8.5    &71      &98      &9  &$>$ 7.0  &\nodata & \\
           &+13.3   &29      &36      &9  &1.8      &\nodata & \\
HD 59975   &+12.8   &70      &72      &7  &6.9      &-3.4    & \\
BD+47 1466 &-4.0    &54      &106     &7  &$>$ 5.4  &\nodata & \\
           &+2.4    &46      &57      &7  &4.5      &\nodata & \\
           &+11.0   &105     &137     &7  &$>$ 10.4 &\nodata & \\
HD 60653   &-4.0    &121     &145     &6  &$>$ 12.0 &-41.1   & \\
AG+47 701  &-1.4    &161     &185     &9  &$>$ 16.0 &\nodata & \\
           &+12.6   &110     &113     &9  &$>$ 10.9 &\nodata & \\
HD 60933   &+3.9    &32      &52      &8  &3.2      &\nodata & \\
           &+11.0   &17      &25      &8  &1.2      &\nodata & \\
           &+14.9   &4       &12      &8  &0.6      &\nodata & \\
BD+47 1470 &+12.1   &64      &82      &6  &$>$ 6.3  &+0.4    & \\
BD+46 1294 &-8.5    &15      &17      &6  &0.9      &\nodata & \\
           &+3.2    &17      &29      &6  &1.4      &\nodata & \\
           &+8.8    &16      &19      &6  &0.9      &\nodata & \\
BD+47 1474 &+9.8    &98      &125     &7  &$>$ 9.7  &-22.8   & \\
BD+47 1476 &+10.3   &116     &148     &8  &$>$ 11.6 &-44.9   & \\

\enddata

\tablenotetext{a}{No obvious interstellar absorption, however narrow stellar line is close to expected velocity}
\tablecomments{Stellar $V_{LSR}$: if the line is too broad or too weak, no velocity is listed}

\end{deluxetable}

\newpage
\onecolumn

\begin{figure}
\figurenum{1}
\plotone{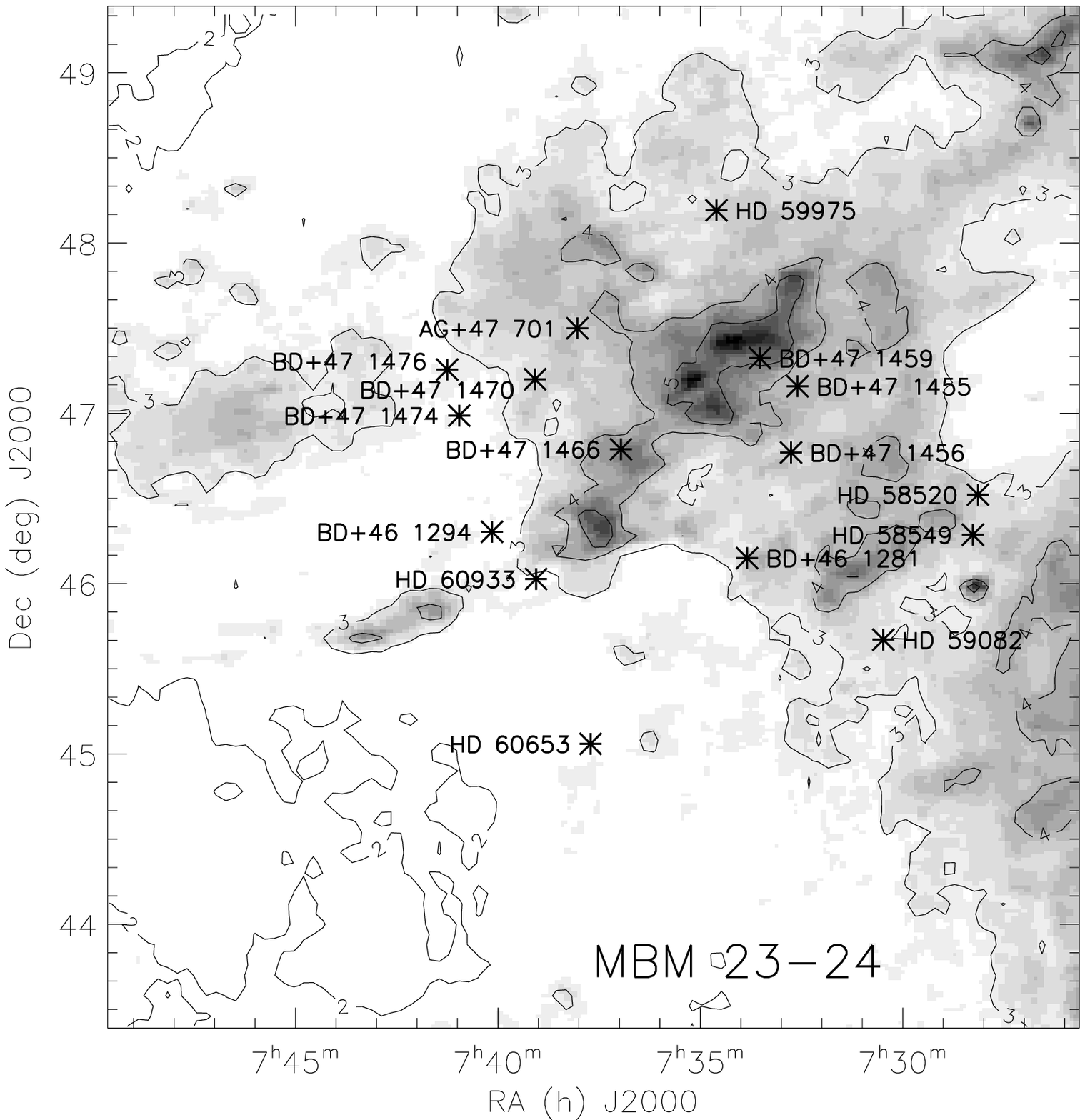}
\figcaption[]{b) MBM~23-24}
\end{figure}

\begin{figure}
\plotone{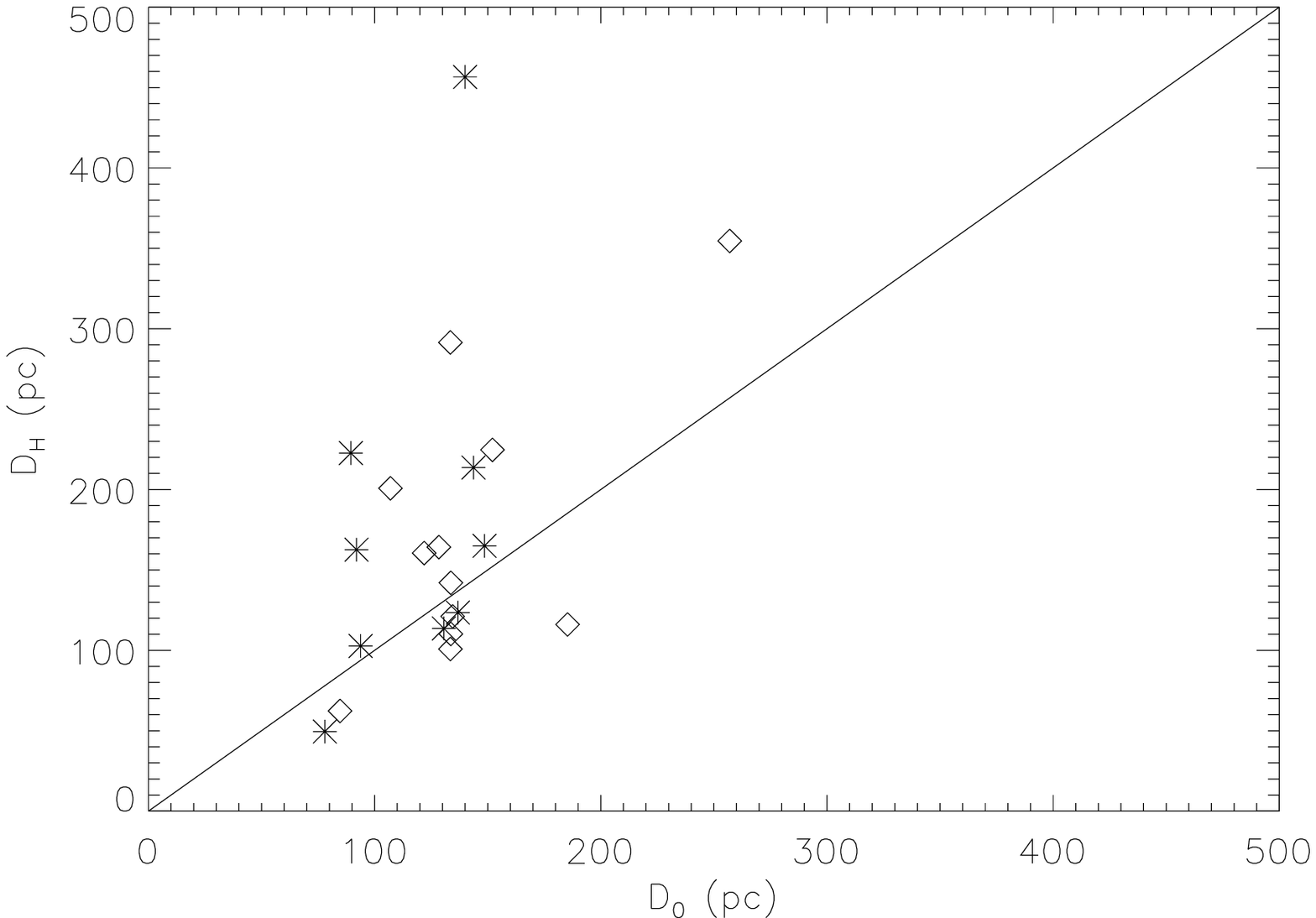}
\figcaption[]{Comparison of spectroscopic, $D_0$, and {\it Hipparcos}, $D_H$, distances.  Starred points do not have published luminosity classes and were assumed to be luminosity class V.  The line is at $D_0 = D_H$.  The effect of interstellar extinction is to decrease $D_0$, while incorrect specification of luminosity class increases $D_0$. \label{fig-dvd}}
\end{figure}

\begin{figure}
\plotfiddle{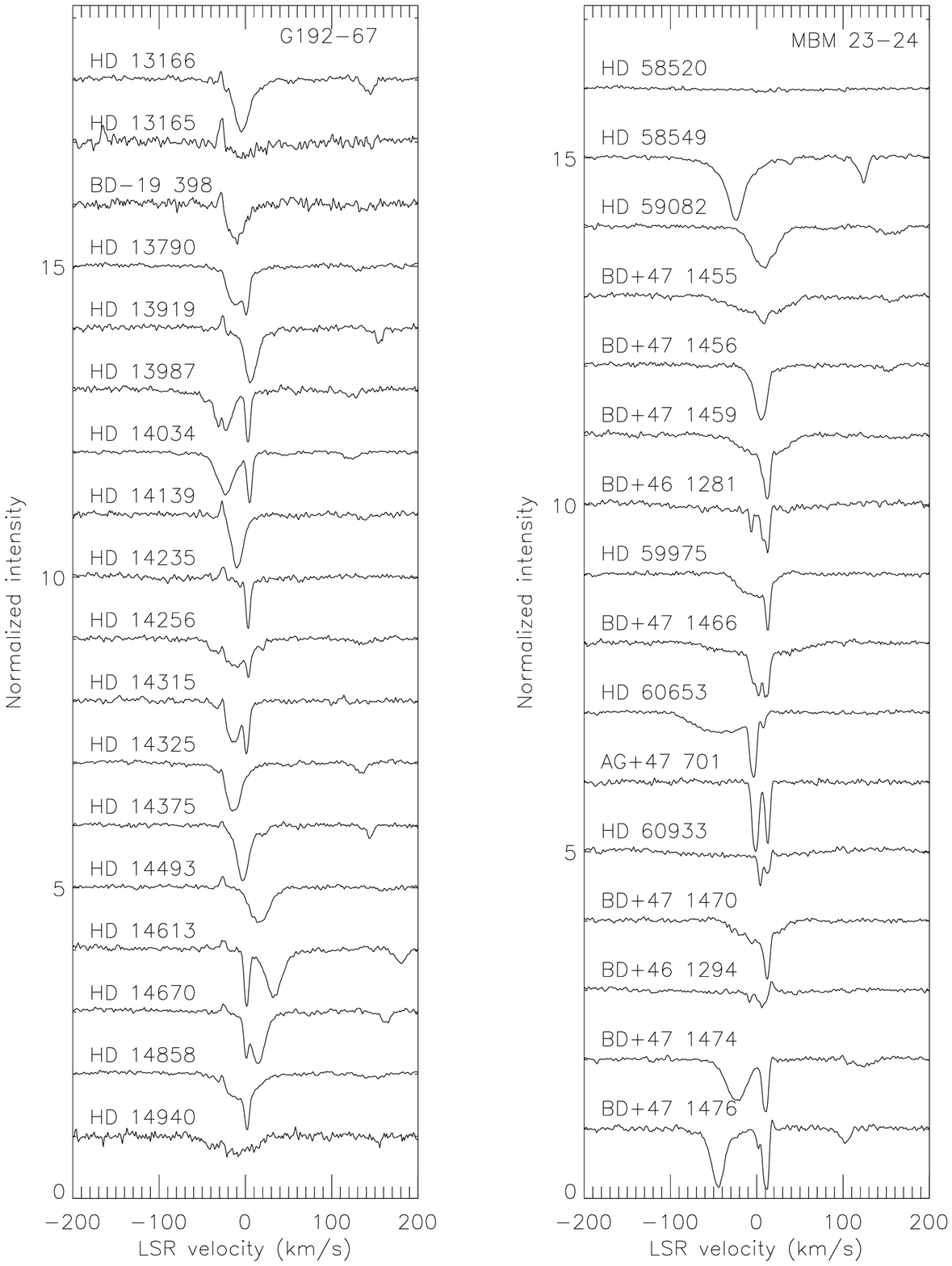}{7.8in}{0}{100}{100}{-260}{0}
\figcaption[]{Spectra of \ion{Na}{1}~D2 lines.  Each spectrum has been normalized by a high-order spline function.  They are displayed such that the continuum level for a spectrum is the zero level for the spectrum above it.  a) G192-67 b) MBM~23-24 \label{fig-spec}}
\end{figure}

\begin{figure}
\plottwo{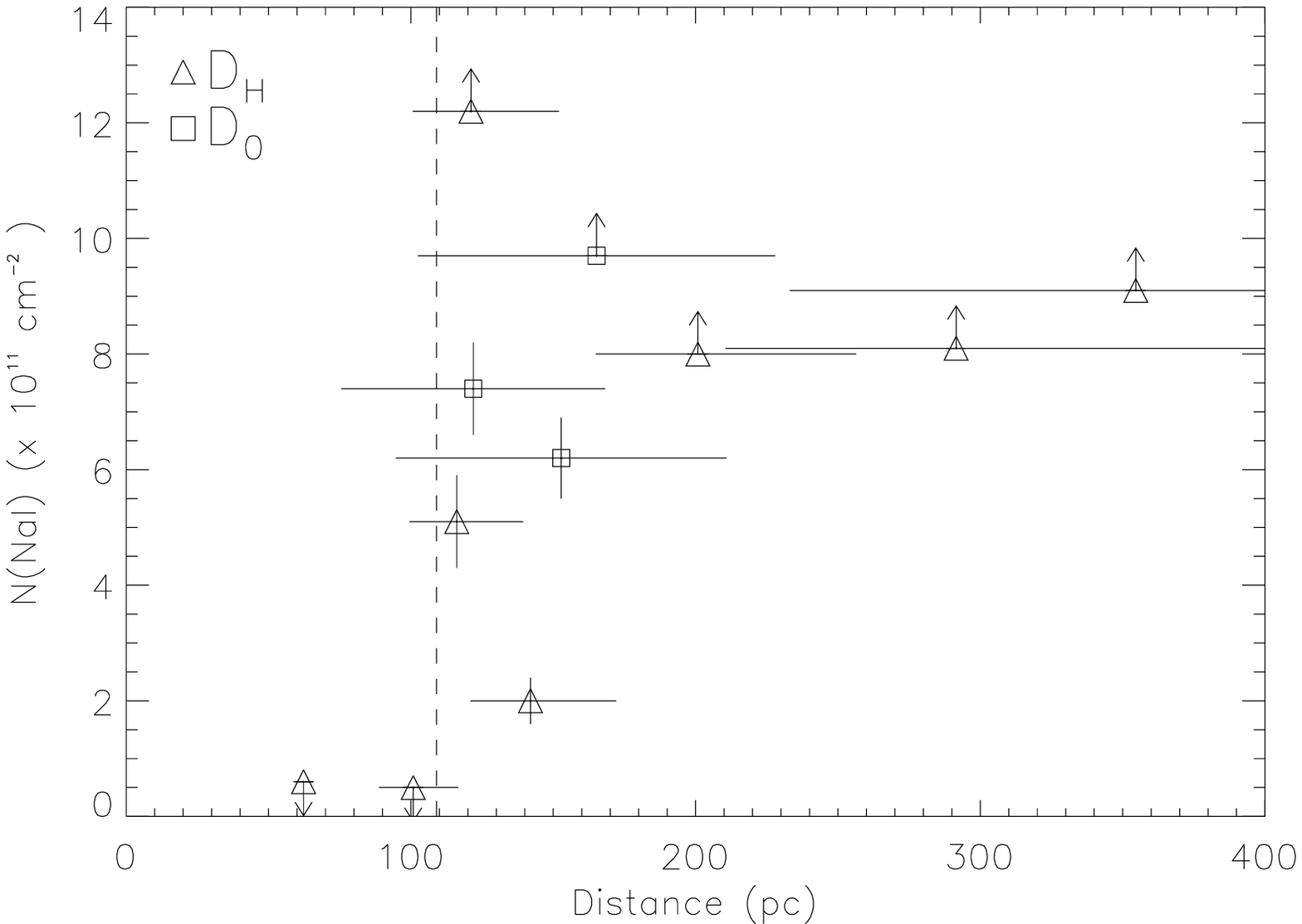}{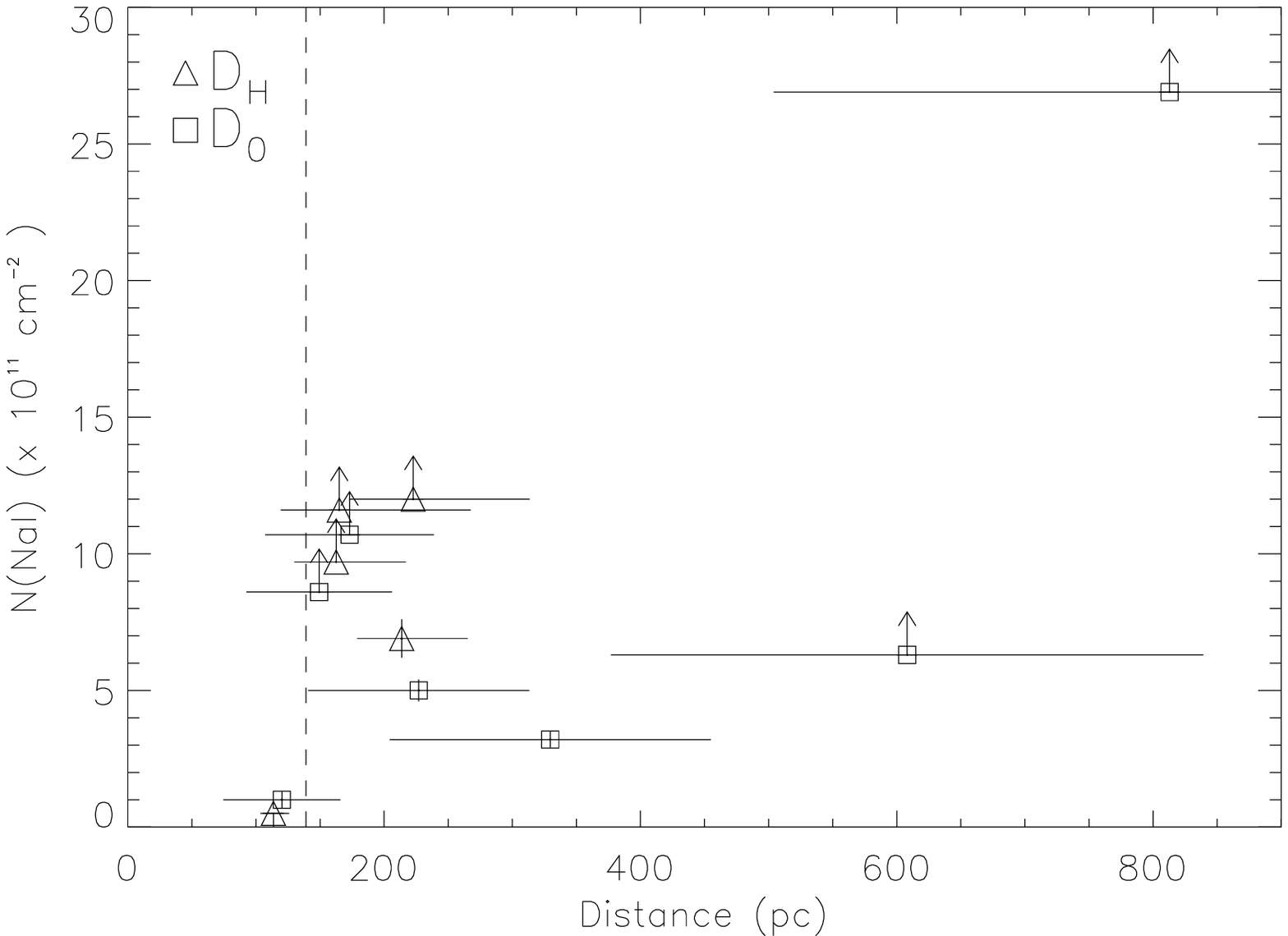}
\figcaption[]{Total column density of the interstellar absorption lines versus stellar distance.  Estimated errors in the {\it Hipparcos} distances are based on the 1-$\sigma$ parallax errors and estimated errors in the spectroscopic distances are set at 38\% as derived in \S 2.1.  The derived distance to the cloud is shown with a dashed line.  a) G192-67 b) MBM 23-24 \label{fig-dist}}
\end{figure}

\end{document}